\documentclass[FBSedit,FBSmath,ecsub]{FBSsuppl}
\usepackage{amsfonts}
\usepackage{amssymb}
\def\b{{\rm b}}

\def\d{{\rm d}}
\def\u{{\rm u}}
\def\p{{\rm p}}
\def\g{{\rm g}}

\def\B{{\rm B}}

\def\L{{\rm L}}
\def\Q{{\rm Q}}
\def\GeV{{\rm GeV}}
\def\MeV{{\rm MeV}}
\def\nb{{\rm nb}}
\def\mb{{\rm mb}}

\title{The Chances to Produce and Detect the $\b\b\bar{\u}\bar{\d}$ Tetraquark at LHC}
\author{D. Janc$^{1}$\thanks{\textit{E-mail address:} 
damijan.janc@ijs.si} M. Rosina$^{1,2}$ D. Treleani$^{3}$ and A. Del Fabbro$^{3}$}
\institute{
{$^1$J.~Stefan Institute, 
              1000 Ljubljana, Slovenia}\\
$^2${Faculty of Mathematics and Physics,
              University of Ljubljana,
              1000 Ljubljana, Slovenia}\\
$^3${Universita di Trieste; Dipartimento di Fisica Teorica, Strada Costiera 11,}\vspace{0.5mm}
{Miramare-Grignano, and INFN, Sezione di Trieste, I-34014 Trieste, Italy.}}

\begin{document}

\maketitle
\begin{abstract}
In the LHC collider a significant rate of events with double parton scattering 
is expected. This will be the leading mechanism for production
of two b$\bar{\textrm{b}}$ pairs. We estimate the probability of binding 
two b quarks into a diquark and the probability of dressing this diquark into
a $\b\b\bar{\u}\bar{\d}$ ($IS^P=01^+$) tetraquark. Calculations shows that
that this bound state of two B mesons is stable against the strong interaction
and has a life time of the order of ps. We estimate that the production rate
at luminosity $\L=0.1$ events per second will be about 6 tetraquarks per hour or more.
\end{abstract}

\section{Introduction}

In the new large hadron collider at CERN on can expect a sizable rate of events
where two heavy b quarks and two heavy $\bar{\b}$ antiquarks are simultaneously
produced. This makes it possible that two heavy quarks or antiquarks combine
to create a double heavy hadron after they dress with light quarks.\\ 
There are two mechanisms for
creation of two heavy quark-antiquark pairs. The first is by a single
parton scattering
where initially only one heavy quark-antiquark pair is produced and the
second pair is then created by the fragmentation of one of
initial quarks. The second mechanism is a double parton scattering.
Calculations show \cite{fab} that the double scattering although power suppressed
gives at the center of the mass energy of 14 TeV a four times larger
cross section for the production of two b-quark pairs. 
For the creation of a heavy diquark one requires that the two quarks are close together 
in momentum space. Since in the fragmentation the two b quarks tend to fly in opposite
directions this also confirms that the dominant mechanism would be two parton scattering.
The creation of the double heavy hadron is then a two step process.
First two heavy quarks which are close together in momentum space
bind into a diquark. We can estimate that the probability for this 
is proportional to the overlap of the wave functions of the quarks which 
are considered to be free with the wave function of the diquark
The second step is then dressing of this diquark with the light quark into
a doubly heavy baryon or dressing with the two light quarks into a 
heavy dimeson.

\section{Phenomenological estimate}

Let me first give a very simple phenomenological estimate
why a BB dimeson should exist with a binding energy of about 100 MeV
and why a similar DD dimeson is unstable against strong interaction \cite{jst}.
In this estimation we consider the heavy dimeson to be a bound
state of two heavy b-quarks combined
in a diquark with the spin 1 and two light antiquarks with the quantum numbers
spin 0, isospin 0.
The starting point is that the light antiquarks around the heavy diquark
behave as in $\Lambda_\b$. This would be exactly true
in the limit where the mass of the b quark goes to infinity and the
heavy diquark is point-like. This correspond to assumption that we 
can neglect the size of the heavy diquark in dimeson.
The second assumption is that the interaction between heavy b quarks
in a colour antitriplet state is half as strong as the interaction
between heavy quark and antiquark in heavy meson. With this assumption 
we can estimate the binding energy of heavy dimeson.  Since the spectroscopy 
of heavy mesons is quite well described with nonrelativistic potential models
we see that the binding energy for a system with half as strong interaction 
$V_{\Q\Q}$ is half the binding energy for a system with 
original interaction $V_{\Q\bar{\Q}}$ and twice lighter particles 
$(m_\Q\to m_\Q/2)$.
$$
\Bigl[\frac{p^2}{m_\Q}+V_{\Q\Q}\Bigr]\psi=\frac{1}{2}\Bigl[
 \frac{p^2}{m_\Q/2}+V_{\Q\bar{\Q}} \Bigr]\psi=\frac{1}{2}E(m_\Q/2)\psi
$$
We obtain $E(m_\Q/2)$ by plotting the binding energy of heavy mesons
as a function of the reduced mass of the system \cite{jst}. We do that for
different choices of constituent quark masses, to estimate the
uncertainty due to quark masses. The plot is very smooth
and we estimate the binding energy of the heavy  bb diquark $E(m_\b/2)/2$ to be
$-390\pm 15$ MeV.
If we combine this with the masses of heavy mesons and $\Lambda_b$ baryon
we can obtain the phenomenological estimate for binding energy of 
the heavy dimeson $\Delta T_{\b\b}$
$$
\Delta T_{\b\b}=m_{\Lambda_\b}+\big[m_{\Upsilon}-E(m_\b)+E(m_b/2)\big]/2-m_\B-m_{\B^*}=
-130\pm 15\, \MeV
$$
while the heavy dimeson with the two c quarks is unbound.
This agrees well with some previous four body calculations \cite{sil} \cite{bri}
in the constituent quark model.

\section{b-quark pair production}

Since at LHC there will be a large parton luminosity we can hope that
the production rate of double heavy  dimesons will be high enough to 
make detection possible.
The dominant contribution to b-quark pairs production will be double
parton scattering - the gluon fusion $(\g\g\to\b\bar{\b})^2$.  
To calculate this cross section one needs to make an assumption
about the two body distribution function $\Gamma$ 
\begin{equation}
\Gamma(x_1,x_2,d)=G(x_1)G(x_2)F(d),
\label{gama}
\end{equation}
where $G(x)$ is the one-body parton function while the parton pair
density is described by the normalized function $F(d)$. The distance of 
the two partons belonging to the same proton is $d$ and the $x_1$ and $x_2$ are
their fractional momenta. Doing so, we neglect the correlations in momenta.
Then the cross section $\sigma_D$ for the double gluon fusion has a form 
\begin{equation}
\sigma_D(\b \bar{\b}\b \bar{\b})=\frac{\sigma_s (\b \bar{\b})^2}{2 \sigma_{eff}},\quad
\sigma_{eff}^{-1}=\int d^2 b F(b)^2,
\label{cs}
\end{equation}
where $\sigma_s(b \bar b)$ is cross section for a fusion of two gluons 
into a b quark-antiquark pair.
For $\sigma_{eff}$ we
take the value measured in the experimental study of double
parton 
collision by CDF at Fermilab in the reaction
$\p\bar{\p}\to\gamma/\pi^0+
3\,\textrm{jets + X}$ \cite{sig}.
 This experimental value is
$14.5\,\mb$. Doing this we assume that also some correlation between quarks
in the same proton are taken into account which were neglected in (\ref{gama}).  
There is also same uncertainty about the single scattering cross section
$\sigma_s$ since the higher order corrections are large and they enter squared 
into the cross section for double scattering (\ref{cs}). For details see 
\cite{fab}. The cross section calculated then from (\ref{cs}), taking into account that
only the pseudorapidity region $1.8<\eta<4.9$ will
be covered by forward LHCb detector, is 85 nb.
For two b quarks to combine into a diquark it is very important that they
are close together in momentum space. Calculations gives that the cross section
is approximately proportional to momentum volume $d\sigma/d^3p\approx 0.4\,\nb/\GeV^3$
up to $p\leq2\GeV$.

\section{Formation of the diquark and dressing}

We assume that the production of the two heavy b quarks in double parton
scattering is uniform inside the protons. Furthermore we take
an instantaneous approximation so that the amplitude $M(p)$ for the formation of the
diquark is simply given by the overlap $M(p)=\langle\phi(p)|\psi\rangle$  of the wave function of the two
b quarks $\phi(p)$ with the wave function of the diquark
$\psi$. This wave function is calculated in constituent quark model used also in 
\cite{sil} and \cite {bri}. So the production cross section is given by
\begin{equation}
\sigma=\int d^3 p \frac{d\sigma}{d^3p}M^2(p)=0.15\,\nb
\label{rez}
\end{equation}  
\\
Dressing of a heavy diquark is governed by the quantum chromodynamics. But
since fragmentation of diquark into a hadron is impossible to calculate using 
perturbative QCD methods, we estimate the probability from experimental data
obtained at Fermilab and at LEP experiment \cite{cdf}. Their results show that the 
probability that the b quark
will be dressed with u or d antiquark is $0.38\pm0.02$, with s antiquark $0.16\pm0.03$ and
the probability for dressing with light quarks to form $\Lambda_b$ is $0.10\pm0.03$.
In our estimation we compared the heavy dimeson with the $\Lambda_\b$ baryon where
the role of diquark in the first case is played by the heavy b quark in the second.
So we assume that
the diquark will get dressed into a dimeson with the same probability as
heavy quark is dressed into a lambda b baryon - that is $0.10\pm0.03$.   
The cross section for the heavy dimeson production is thus
$$
\sigma=0.015\,\nb
$$
which will give at the expected luminosity the production rate of 5-6 events
per hour.

\section{Decay and detection}

We expect that the $\b\b\bar{\u}\bar{\d}$ tetraquark will be
stable against strong and electromagnetic decay and will decay only weakly.
The main channel would be the decay of one of the b quarks into c quark followed 
by the weak decay of the second b quark. But this channel will be difficult to
distinguish from the channel of two unbound B mesons which will be 
the main background contribution. The more characteristic decay channel would be the 
the formation of the $\Upsilon$ and light meson through $\b \to \bar{\b}$ oscillation
inspired by $\B^0\to\bar{\B}^0$ oscillations. But this oscillations are
here negligible since the tetraquark is not degenerate with final 
state of two unbound mesons.
We are still searching for new ideas for the detection of doubly b tetraquarks.

\end{document}